\documentclass[preprint2]{aastex1}
\usepackage{here}
\begin{document}

\title{THE ENIGMA OF THE OPEN CLUSTER M29 (NGC 6913) SOLVED}

\author{V. Strai\v{z}ys\altaffilmark{1}, K.
Mila\v{s}ius\altaffilmark{1},
R. P. Boyle\altaffilmark{2}, F. J. Vrba\altaffilmark{3},
U. Munari\altaffilmark{4},
N. R. Walborn\altaffilmark{5},\\ K. \v{C}ernis\altaffilmark{1},
A. Kazlauskas\altaffilmark{1}, K. Zdanavi\v{c}ius\altaffilmark{1},
R. Janusz\altaffilmark{6}, J. Zdanavi\v{c}ius\altaffilmark{1} and
V. Laugalys\altaffilmark{1}}

\altaffiltext{1}{Institute of Theoretical Physics and Astronomy, Vilnius
University, Go\v{s}tauto 12, Vilnius LT-01108, Lithuania}
\altaffiltext{2}{Vatican Observatory Research Group, Steward Observatory,
Tucson, AZ 85721, U.S.A.}
\altaffiltext{3}{U.S. Naval Observatory Flagstaff Station, P.O. Box 1149,
Flagstaff, AZ 86002, U.S.A.}
\altaffiltext{4}{INAF Astronomical Observatory of Padova, I-36012,
Asiago (VI), Italy}
\altaffiltext{5}{Space Telescope Science Institute,
3700 San Martin Drive, Baltimore, MD 21218, U.S.A. Operated by
AURA, Inc., under NASA contract NAS5-26555}
\altaffiltext{6}{University School `Ignatianum', Cracow, Poland}

\shorttitle{THE ENIGMA OF THE OPEN CLUSTER M29 SOLVED}
\shortauthors{V. Strai\v{z}ys et al.}


\begin{abstract}

Determining the distance to the open cluster M29 (NGC 6913) has proven
difficult, with distances determined by various authors differing by a
factor of two or more.  To solve this problem, we have initiated a new
photometric investigation of the cluster in the Vilnius seven-color
photometric system supplementing it with available data in the {\it BV}
and {\it JHK}$_s$ photometric systems, and spectra of the nine brightest
stars of spectral classes O and B. Photometric spectral classes and
luminosities of 260 stars in a 15$\arcmin$\,$\times$\,15$\arcmin$ area
down to $V$ = 19 mag are used to investigate the interstellar extinction
run with distance and to estimate the distance of the Great Cygnus Rift,
$\sim$\,800 pc.  The interstellar reddening law in the optical and
near-infrared regions is found to be close to normal, with the ratio of
extinction to color excess $R_{BV}$ = 2.87.  The extinction $A_V$ of
cluster members is between 2.5 to 3.8 mag, with a mean value of 2.97 mag
or $E_{B-V}$ = 1.03.  The average distance of eight stars of spectral
types O9--B2 is 1.54\,$\pm$\,0.15 kpc.  Two stars from the seven
brightest ones are field stars:  HDE 229238 is a background B0.5
supergiant and HD 194378 is a foreground F star.  In the intrinsic
color-magnitude diagram, seven fainter stars of spectral classes B3--B8
are identified as possible members of the cluster.  The 15 selected
members of the cluster of spectral classes O9--B8 plotted on the $\log
L/L_{\odot}$ vs.  $\log T_{\rm eff}$ diagram, together with the
isochrones from the Padova database, give the age of the cluster as
5\,$\pm$\,1 Myr. \end{abstract}

\keywords{stars:  fundamental parameters, classification
--- Galaxy:  open clusters and associations:  individual (M29, NGC 6913,
Cyg OB1)}

\section{INTRODUCTION}

Despite its impressive appearance, similar to a tiny Pleiades, the
cluster M29 (NGC 6913) is a difficult object to understand.  A glance at
a deep exposure in blue or red filters shows that it is located in a
small more transparent bay of a huge system of dark clouds known as the
Great Cygnus Rift.  The most prominent in the cluster are seven stars
with $V$ magnitudes 8.6--10.2, located within 5 arcmin:  HD 194378, HDE
229221, HDE 229227, HDE 229234, HDE 229238, HDE 229239 and BD+38 4067.
All of them are of O9--B0 spectral classes, except for HD 194378 which
is a known foreground F star.

The first evaluations of the cluster members and their distances were
quite uncertain due to the low accuracy of photographic photometry, the
lack of spectral and luminosity classifications and of insufficient
calibration accuracy of the photometric data
\citep{Zug1933,Becker1948,Tifft1958}.  Tifft concluded that the cluster
looked like it consisted of two groups at different distances -- 1.6 and
2.1 kpc.  The brightest M29 stars were first classified in the MK system
by \citet{Roman1951, Morgan1953,Morgan1955,Morgan1956} and
\citet{Hiltner1956}.

The next study of M29 was done by \citet{Hoag1961} combining
photoelectric and photographic {\it UBV} photometry.  For the distance
determination a $V_0-M_V$ versus $V_0$ diagram for B-type stars was used
\citep{Johnson1960}.  Intrinsic $(B-V)_0$ values were found from the
reddening-free parameters $Q_{UBV}$, thus spectral types were not used.
The mean extinction $A_V$ = 3.21 mag and a distance of 1.15 kpc were
found by \citet{Johnson1961}.  However, no explanations were given for
how the cluster members were selected, and no mention of the group of
the brightest stars of spectral types O9--B0 was given.  It seems that
these stars were ignored.

A few years later, \citet{Hoag1965} and \citet{Walker1968} obtained
H$\gamma$ line photoelectric photometry of the brightest stars of M29
calibrated in absolute magnitudes.  The distances to these stars were
found to be much smaller than their spectroscopic distances.
\citet{Crawford1977} using {\it uvby}H$\beta$ photometry found a similar
result:  the distances to the brightest stars calculated with the
absolute magnitudes $M_V$, obtained from calibration of H$\beta$
absorption, were half of their spectroscopic distances.  All this means
that for the brightest M29 stars something was wrong either with their
MK luminosity classes, or with the calibration of the hydrogen line
strengths in luminosities.

Two subsequent investigations were aimed at increasing the number of
stars in the vicinity of M29 with {\it UBV} photometry and MK spectral
types.  \citet{Joshi1983} have observed in {\it UBV} 103 stars covering
the 40$\arcmin$\,$\times$\,40$\arcmin$ area for which the cluster
membership probabilities from proper motions were estimated by
\citet{Sanders1973}.  Later on, most of these stars were classified in
the MK system by \citet{Wang2000} from low-resolution spectra.

The first $M_{\rm bol}$ versus $\log T_{\rm eff}$ HR diagram of M29 was
obtained by \citet{Massey1995}.  Since the MK types were known only for
the five brightest stars, other B and A stars were classified
approximately by their $Q_{UBV}$ parameters.  The age of the cluster
from the comparison with isochrones, 4--6 Myr, mostly rests on the five
brightest stars accepting their distance at 2.2 kpc.  Later on,
\citet{Liu1989} and \citet{Boeche2004} found that most of these stars
are spectroscopic binaries exhibiting periodic variations of radial
velocities.

From our review of photometric and spectroscopic investigations of M29
we conclude that the previous selection of cluster members is in
question.  Looking along the Local spiral arm, the early-type stars can
be found at all possible distances, up to the edge of the arm at 4--5
kpc.  Thus, the spectral class of a star without additional information
about its distance is of little value for membership estimation.  The MK
typing of stars should be helpful but, unfortunately, discrepancies
among luminosity classes of O9--B0 stars are substantial.  The
reddening-free parameter $Q_{UBV}$ applied in some studies to identify
B-type stars does not give luminosities, thus it is useless in
estimating the membership.  Also, a B-type star sample, selected in this
way, can be contaminated by A and F stars.  Proper motions of stars,
used in some investigations as the membership criteria, in the direction
along the Local spiral arm are too small to be useful at large
distances.  All these considerations lead to the conclusion that until
now the cluster membership is the main problem which prevents
determining an accurate distance to M29 and its age.

With the aim to identify more early-type stars in the direction of M29
and to get more confident information about their luminosities and
distances, we decided to undertake a new investigation of the cluster in
the Vilnius seven-color photometric system which allows classification
of most types of stars in spectral class and luminosity in the presence
of variable interstellar reddening.  Unfortunately, the system is not
sufficiently sensitive to luminosities for the stars of spectral classes
O--B0, thus we are not able to verify luminosity classes of the
brightest stars of M29. Therefore we decided to obtain new medium
resolution spectra of these stars in order to verify their luminosity
classes.

\section{SEVEN-COLOR PHOTOMETRY AND\\ CLASSIFICATION}

The results of CCD photometry in the Vilnius seven-color system,
published by \citet[][hereafter Paper I]{Milasius2013}, cover a 1.5
square degree area centered on the cluster at RA (J2000) = 20$^{\rm
h}$\,24.0$^{\rm m}$, DEC (J2000) = +38$\degr$\,30$\arcmin$.  Magnitudes
and colors were determined for 1752 stars down to $V$ = 19.5 mag.
However, the limiting magnitudes $V$ are not uniform in the whole area
since three telescopes of different apertures and field sizes were used.
The faintest stars have been measured in the central
13$\arcmin$\,$\times$\,13$\arcmin$ area observed with the 1.8 m VATT
telescope on Mt. Graham.  A larger circular area with a 22$\arcmin$
diameter was observed with the 1 m telescope at the USNO Flagstaff
Station down to a limiting magnitude of 17 mag.  The largest 1.25$\degr$
square area was observed with the 35/51 cm Maksutov-type telescope of
the Mol\.etai Observatory in Lithuania with a limiting magnitude of 15
mag.  The magnitudes and colors of stars observed with the two or three
telescopes were averaged.  Standard stars used for the present
photometry were observed in the Vilnius system photoelectrically by
\citet{Kazlauskas1986}.

The photometric data were used to classify about 70\,\% of stars in
spectral and luminosity classes and peculiarity types.  The results of
the classification are given in Paper I.

For this investigation of the cluster we used the stars only in the
central 15$\arcmin$\,$\times$\,15$\arcmin$ area (Figure 1) which have
been measured and classified to the faintest limiting magnitude.


\begin{figure}[!t]
\includegraphics[width=75mm]{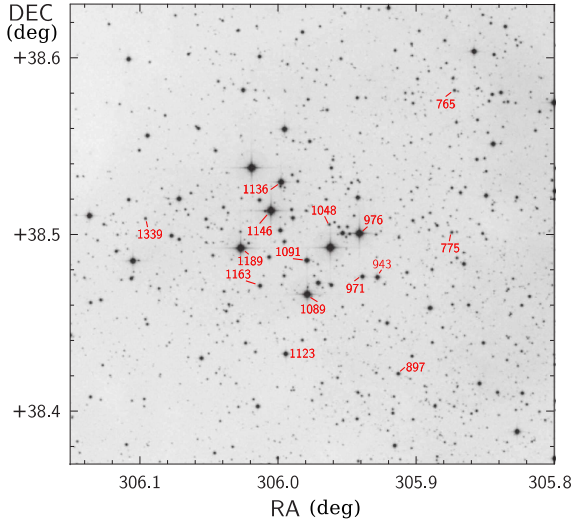}
\caption{The 15$\arcmin$\,$\times$\,15$\arcmin$
field around the M29 cluster investigated in this study.  The
background image is the DSS2 Red image from SkyView. The numbered stars
are 15 probable cluster members from Table 2. The numbers
are from the catalog published in Paper I.\label{fig1}}
\end{figure}


\begin{figure*}[!t]
\centerline{\includegraphics[width=140mm]{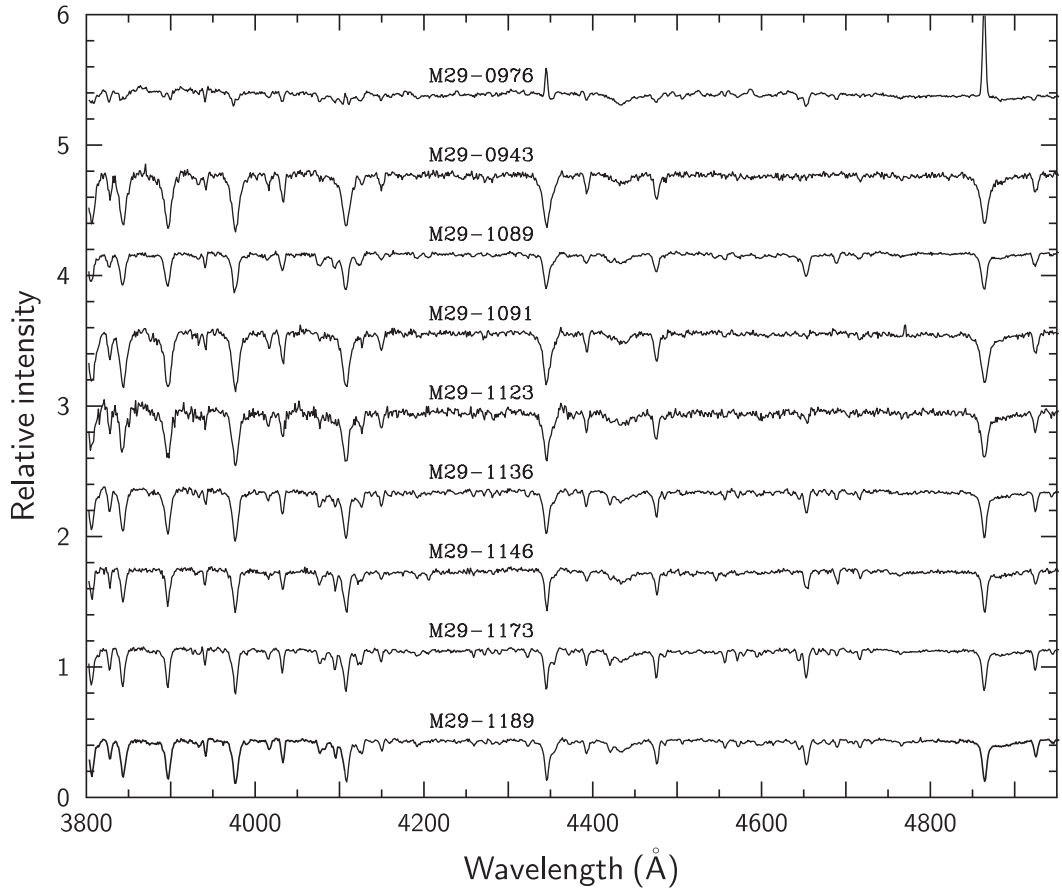}}
\caption{The spectra of nine M29 stars of spectral classes O and B
obtained on the 1.22 m telescope of the Asiago Observatory.  The numbers
of the stars indicated correspond to the catalog by Mila\v{s}ius et al.
(2013). The fits files of the spectra are available in the
electronic edition of AJ.\label{fig2}}
\end{figure*}

\def\tablerule{\noalign{\vskip.7ex}\noalign{\hrule}\noalign{\vskip.6ex}}


\begin{table*}
\footnotesize
\caption{Spectral types of the nine brightest O9--B2 stars of M29.}
\begin{tabular}{rrlrrr}
\tablerule
{Paper I }         & {HD, BD }  &{Spectral type [Reference] }& { $V$ } &{$A_V$ } &{$d$ kpc}\\
\tablerule
943  &  --      & B2\,V, SB [5]; B2\,V [6]                                                       & 11.77 & 2.83 & 1.34 \\
976  & 229221   & B0\,II:e [1], B0pe [2], Be [3], B0.2\,IIIe [4], B0\,IIIe [5], B0:pe [6]        & 9.22 & 3.41 & 1.45 \\
1089 & 229227   & B0\,III [1], B0\,II [2], B0.2\,III [3], O9.7\,III [4], B0\,V [5], O9.7\,III(n) [6] & 9.34 & 3.06 & 1.80 \\
1091 & --       & B3\,V [6]                                                                      & 11.41 & 3.10 & 1.52 \\
1123 & --       & B1-1.5\,V [6]                                                                  & 11.91 & 3.83 & 1.64 \\
1136 & +38 4067 & B0.2\,III [3], B0\,Ib/II [5], B0.5-0.7\,V [6]                                  & 10.20 & 3.00 & 1.55 \\
1146 & 229234   & O9\,II [1], O9.5\,III [2], O9\,If [3], O9\,II [4], O9\,Ib, SB [5], O9\,III [6] & 8.91  & 3.03 & 1.64 \\
1173 & 229238   & B0.5\,II [1], B0.5\,Ib [2], B0\,I [3], B0.2\,II [4], B0\,I/II, SB [5], B0.5\,Ia [6] & 8.86 & 3.10 & 3.56 \\
1189 & 229239   & B0.5\,IIe [1], B1\,Iab [2], B0.5\,IV [3], B0.2\,III [4], B0\,I, SB [5], B0.2\,III [6] & 8.90 & 3.26 & 1.34 \\
\tablerule
&&&&&\\
\noalign{\parbox{15cm}{References --- [1] \citet{Roman1951}; [2] \citet{Morgan1953, Morgan1955}; [3] \citet{Massey1995};
[4] \citet{Negueruela2004}; [5] \citet{Boeche2004}; [6] This paper.}}
\end{tabular}
\vskip-5mm
\end{table*}

The spectra of the nine brightest stars (Figure 2) were obtained with
the 1.22 m telescope at the Asiago Observatory, equipped with a B\&C
spectrograph and a 600 lines/mm grating providing a dispersion of 1.17
\AA/pix and a resolving power of FWHM (PSF) = 2.1 pix.  The detector was
an ANDOR iDus DU440A CCD with the 2048\,$\times$\,512 array and 13.5
$\mu$m square pixels.  The recorded wavelength range covers 3487--5885
\AA.  The nearby standard HR 8004 was used for the flux calibration.
Data reduction in IRAF followed strictly the procedures in
\citet{Zwitter2000}, and involved all classical steps of correction for
bias, dark, flat, sky background and instrumental response.  Spectral
classification in the MK system was carried out by one of us (NRW) using
the standards from \citet{Walborn1990} and \citet{Sota2011}.  The
classification results with comments are given below and in Table 1
which also gives the results of spectral classifications of other
authors, based only on the slit spectra with the resolution comparable
or higher than that used for establishing the MK system, 125 \AA/mm at
H$\gamma$ \citep{Morgan1943}.

{\bf Star 1089}:  O9.7 III(n).  The ratio He\,II $\lambda$4541 / Si\,III
$\lambda$4552 $\sim$\,1 defines the spectral class.  The ratio He\,II
$\lambda$4686 / He\,I $\lambda$4713 defines luminosity class, in
reasonable agreement with Si\,IV / He\,I near H$\delta$ which have broad
lines.  The $\lambda$4686 / $\lambda$4713 ratio $\sim$\,1 would
correspond to luminosity class II at this type, whereas it should be
larger than it is for luminosity V. This ratio changes very rapidly
between spectral classes late O and B0, no doubt contributing to
discrepancies in previous classifications.  The broadening parameter is
approximate, in the absence of standards.

{\bf Star 1136}:  B0.5--0.7\,V.  The ratio Si\,IV $\lambda$4089 $\geq$
Si\,III $\lambda$4552 for spectral class.  Luminosity class is based on
He\,II $\lambda$4686 $\sim$ He\,I $\lambda$4713 at such a late type.
O\,II and C\,III would favor a brighter luminosity class, but the helium
line at $\lambda$4686 disallows this.  Of course, if composite (a
possible interpretation), the classification is a compromise.

{\bf Star 1146}:  O9\,III.  The ratios He\,II $\lambda$4541 $\sim$ He\,I
$\lambda$4387 and He\,II $\lambda$4200 $\sim$ He\,I $\lambda$4144 define
the spectral class.  The ratio He\,II $\lambda$4686 / He\,I
$\lambda$4713 defines luminosity class; the same remarks as for star
1089 above (except for good line quality here).

{\bf Star 1173}:  B0.5\,Ia.  Good match with $\kappa$ Ori in all
criteria.

{\bf Star 1189}:  B0.2\,III.  Criteria the same as above.  Again, the
fine divisions in spectral class are important, because if they are off,
the luminosity class will be wrong since those criteria also evolve with
the spectral classes.

The rest of the stars are classified with less confidence:

{\bf Star 943}: B2\,V.

{\bf Star 976}:  B0:pe.  The absence of clear $\lambda$4200 and
$\lambda$4541 eliminates earlier types, while C\,III $\lambda$4650 and
He\,II $\lambda$4686 disfavor later ones.  Luminosity classification is
not realistic due to emission lines.

{\bf Star 1091}: B2\,V.

{\bf Star 1123}: B1-1.5 V.

Luminosity classification at late-O and early-B types is difficult,
because the luminosity criteria change very rapidly with spectral type,
so that small errors (or uncertainties) in the latter entail large
errors in the former and in the derived absolute magnitudes.  New
interpolated spectral types have been introduced to alleviate this
problem \citep{Sota2011, Sota2014}.  Also, there are some spectra with
discrepancies between the Si\,IV / He\,I and He\,II / He\,I luminosity
criteria, the resolution of which becomes arbitrary.  Some of these
problems are no doubt related to the endemic multiplicity of massive
stars, in terms of both composite spectra and binary evolution.  These
issues likely contribute to discrepancies among different classifiers.
They are extensively discussed by \citet{Walborn2014}.

\section{INTERSTELLAR REDDENING LAW}


\begin{figure}[!t]
\centerline{\includegraphics[width=78mm]{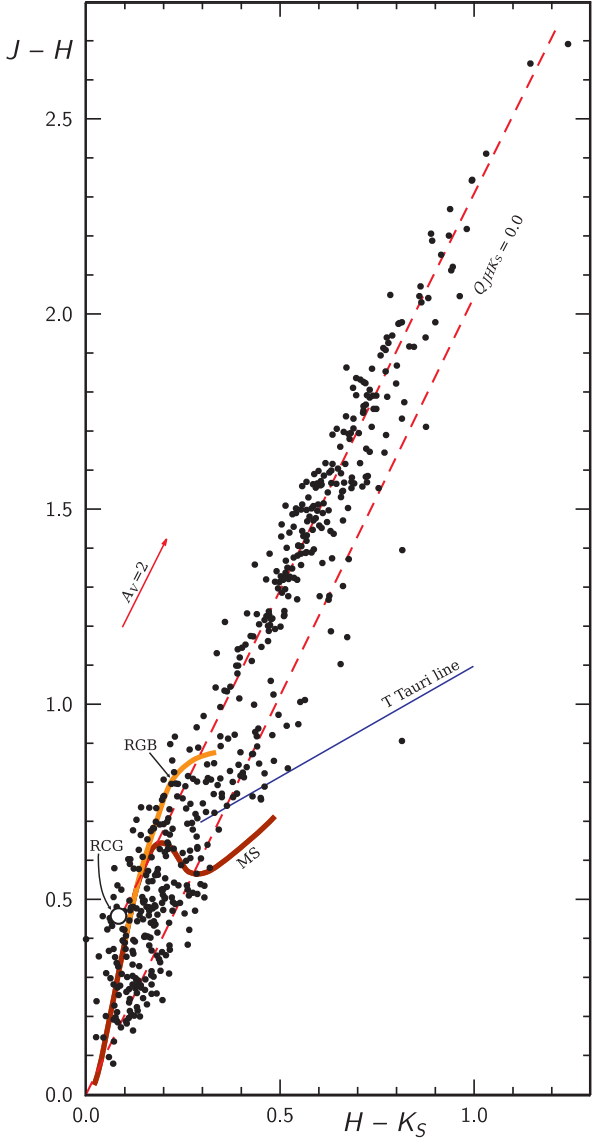}}
\caption{The $J$--$H$ vs. $H$--$K_s$ diagram for 2 MASS stars in the
15$\arcmin$\,$\times$\,15$\arcmin$ box centered on M29. The main
sequence (MS, brown belt), red giant branch (RGB, orange belt), the
intrinsic locus of red clump giants (RCG) with its reddening line, the
reddening line corresponding to $Q_{JHK_s}$ = 0, and the intrinsic T
Tauri line are shown.\label{fig3}}
\end{figure}

For the photometric classification of stars in the Vilnius system by
interstellar reddening-free $Q$-parameters, the normal interstellar
extinction law \citep[see][]{Straizys1992} was applied. Since in the
literature some authors discuss possible peculiarities of the reddening
law in Cygnus, we have decided to verify the law by determining the
slope of the interstellar reddening line in the $J$--$H$ vs. $H$--$K_s$
diagram of the 2MASS system \citep{Skrutskie2006} and the ratios of
color excesses $E_{V-m} / E_{B-V}$ where $m$ are the near-infrared
magnitudes $J$, $H$ and $K$.

In the 15$\arcmin$\,$\times$\,15$\arcmin$ area centered on M29 we
selected 532 stars with $J$, $H$, $K_s$ accuracies of $\leq$\,0.05 mag.
In Figure 3 we plot the $J$--$H$ versus $H$--$K_s$ colors for these
stars, the intrinsic positions of the main sequence, red giant branch
and red clump giants (RCGs, central helium-burning stars) from
\citet{StraiLaz2009}.  Also, the intrinsic line of the classical T Tauri
stars from \citet{Meyer1997} and the reddening line for $Q_{JHK_s}$ =
0.0 are shown.  The classical T Tauri stars usually are located within
the angle formed by these two lines.  The reddened RCGs, together with
normal (hydrogen-burning) red giants of spectral classes G5 to M5, form
a common belt with a width of $\Delta$\,$H$--$K_s$ = 0.15 mag, see
\citet{Straizys2009}.  However, in this belt the RCGs should dominate
since they outnumber by a factor of 10 the space density of normal
giants \citep{Perryman1995, Perryman1997, Alves2000}.  

In Figure 3, the reddening line of RCGs (with a certain amount of normal
giants) can be drawn through their intrinsic position (the large open
circle) and the middle of the belt above $J$--$H$ = 1.0 (the broken
line).  We did not take stars in the belt lower than $J$--$H$ = 1.0
since they are contaminated by stars of other luminosities with small
reddenings.  The slope of this line for 210 stars is $E_{J-H} /
E_{H-K_s}$ = 2.044$\pm$\,0.080.  This slope is quite close to the normal
value 2.0 found in other galactic longitudes \citep{Straizys2008b,
Straizys2008a}.  Thus, we will consider that in the
direction of M29 the interstellar reddening law in the near infrared
wavelengths is close to normal.

Another test of the extinction law in the investigated area is the
application of the ratio of color excesses $E_{V-J}/E_{B-V}$,
$E_{V-H}/E_{B-V}$ and $E_{V-K}/E_{B-V}$ with the $J$, $H$ and $K$
magnitudes from 2MASS.  For this aim we used a list of B and A stars
identified in the central 15$\arcmin$\,$\times$\,15$\arcmin$ area.
Their $V$ and $B$--$V$ data were taken from \citet{Massey1995} and the
AAVSO APASS survey, DR7\footnote{~http://www.aavso.org/apass}, and the
near-infrared magnitudes from 2MASS.  The ratios of color excesses were
calculated separately for the \citet{Massey1995} and the APASS data and
then transformed to the ratios $R_{BV}$ = $A_V$ / $E_{B-V}$ with the
equations from \citet{Fitzpatrick1999}.  The average value of $R_{BV}$
is found to be 2.87 $\pm$ 0.16, i.e., it is somewhat lower than a normal
value of 3.15.  This lower value corresponds to $R_{YV} = A_V / E_{Y-V}$
= 3.83 in the Vilnius system (the normal value is 4.16).

According to \citet{Fitzpatrick2007}, there is no unique relation
between the ratio $R_{BV}$ and the slope of the ultraviolet part of the
extinction law.  Thus, the lower value of $R_{BV}$ found in the area
does not mean that the ratio of ultraviolet to visual color indices (for
example $E_{U-B} / E_{B-V}$) should be also abnormal.  To verify this,
for photometric classification of stars in the area with interstellar
reddening-free $Q$-parameters we have applied the ratios of color
excesses corresponding both to the normal and the so-called `Cygnus'
law, with a somewhat smaller change of slope near the `knee' of
interstellar extinction law at 4350 \AA.  The classification errors of
individual stars were found to be lower in the case of the normal law.

\section{INTERSTELLAR EXTINCTION}

The 1.5 square degree area investigated in Paper I exhibits a very
uneven density distribution of faint background stars since in this
direction the line of sight is close to the edge of the Great Cygnus
Rift dust clouds.  The left side of the area is covered by the Rift
clouds, and the right upper (northwestern) part of the area is the most
transparent.  The central area around the cluster and the lower part of
the area exhibit an intermediate (although quite variable) obscuration.
In the present paper we investigated the extinction only in the
15$\arcmin$\,$\times$\,15$\arcmin$ area centered on the cluster.

For the stars classified in the MK system, color excesses,
extinctions and distances were calculated with the equations
\begin{equation}
E_{Y-V} = (Y-V)_{\rm obs} - (Y-V)_0,
\end{equation}

\begin{equation}
A_V = 3.83 E_{Y-V},
\end{equation}

\begin{equation}
\log d = 0.2 (V - M_V + 5 - A_V),
\end{equation}
where $(Y-V)_0$ and $M_V$ are the intrinsic color indices and absolute
magnitudes for the corresponding MK types from \citet{Straizys1992},
with a correction of --0.1 mag to absolute magnitudes, adjusting their
scale to the new distance modulus of the Hyades, $V-M_V$ = 3.3
\citep{Perryman1998}.  To calculate color excesses for O--B3 stars,
spectroscopic MK types from the present paper were used (see Table 1 in
section 2).  For other stars photometric spectral and luminosity
classes were applied.


\begin{figure}[!t]
\centerline{\includegraphics[width=75mm]{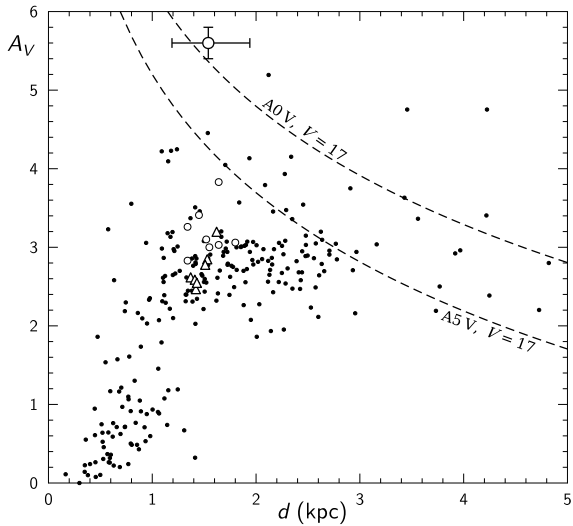}}
\caption{Extinction vs. distance diagram for the stars classified in MK
spectral types in the 15$\arcmin$\,$\times$\,15$\arcmin$ area centered
on the cluster.  The probable cluster members classified
spectroscopically are shown as open circles, and those classified
photometrically as triangles.  The two curves show the limiting
magnitude effect for A0\,V and A5\,V stars at $V$ = 17 mag.  The cross
of maximum (3$\sigma$) errors is shown at the distance of the cluster
(1.54 kpc).\label{fig4}}
\end{figure}

In Figure 4 we show the plot of the extinction $A_V$ vs. distance $d$ in
kpc for 260 stars in the 15$\arcmin$\,$\times$\,15$\arcmin$ area
classified in the Vilnius system.  The two curves in the upper right
part of the diagram exhibit the limiting magnitude effect for A0\,V and
A5\,V stars at $V$ = 17 mag.  We accepted this limiting magnitude
because for fainter stars the number of stars with two-dimensional
classification sharply decreases.  These curves explain why the stars
with high reddenings at large distances are absent.  For example, above
the curve corresponding to A0\,V stars, only a few luminosity IV stars
fainter than $V$ = 17 mag have been observed and classified.  The
maximum (3$\sigma$) error cross is plotted at $d$ = 1.54 kpc, the
distance to the cluster which will be estimated in the next section.  It
corresponds to the distance errors for $\Delta M_V$ = $\pm$\,0.5 and the
extinction errors $\pm$\,0.2 mag.

The distribution of stars in Figure 4 shows a steep increase of
extinction at about 500--600 pc to a value of $\sim$\,3.5 mag.  No
doubt, this jump is related to the cloud system of the Great Cygnus
Rift.  In reality, the clouds are more distant than this rise of the
extinction, because of the negative distance errors of stars and the
presence of stars with the unresolved binarity.  A typical
error of absolute magnitude $M_V$ for B, A and F stars of luminosity
classes V--III, corresponding to one luminosity class, is about
$\pm$\,0.5, and this makes scatter of the calculated distances of stars
to both sides by a factor of 1.26.  This means that the distances of
reddened stars, located apparently at 500 pc, must be multiplied by 1.26
to get the real distance of the cloud, 630 pc.

Another effect, which causes the shortward scatter of the apparent
distances, is the unresolved binarity.  If both components of such a
binary are of the same luminosity, the real distance of the binary
should be larger by a factor of 1.41.  This shifts the stars from 500 pc
to 705 pc.  If the luminosity difference of the components is larger,
the expected shift will be smaller.  If the unresolved binary has an
additional shift shortwards due to the negative luminosity error, the
real distance to the front side of the cloud must be increased by a
factor of 1.26\,$\times$\,1.41 = 1.78.  In this case the real distance
of the front side of the cloud is 500\,$\times$\,1.78 = 890 pc.  The
conclusion is that the front side of the cloud is somewhere between 700
and 900 pc.


\begin{figure}[!t]
\includegraphics[width=75mm]{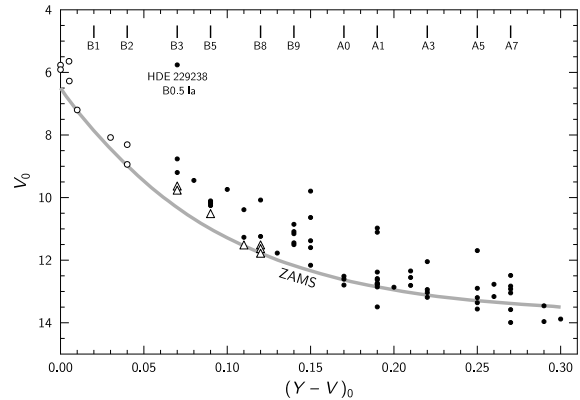}
\caption{Intrinsic color-magnitude diagram for the O-B-A stars in the
15$\arcmin$\,$\times$\,15$\arcmin$ box in the Vilnius system.  The
probable members of M29 are shown as open circles (classified
spectroscopically) and triangles (classified photometrically).  The grey
line is the ZAMS for a distance modulus of $V_0$--$M_V$ = 10.94.  The
spectral classes for luminosity V-III stars corresponding to the
intrinsic $Y$--$V$ values are shown at the upper axis.\label{fig5}}
\end{figure}


\begin{figure}[!t]
\includegraphics[width=75mm]{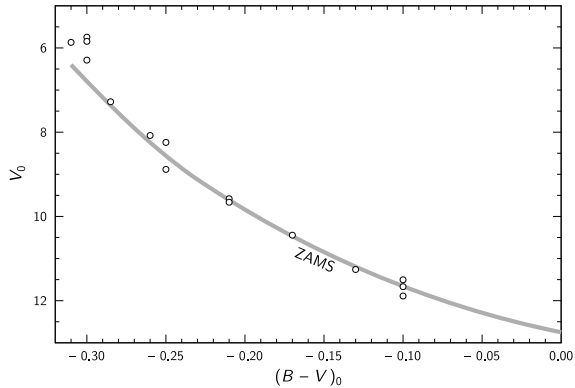}
\caption{Intrinsic color-magnitude diagram in the {\it BV} system
for the 15 probable cluster members of spectral classes O9--B8. The grey
line is the ZAMS for a distance modulus $V_0$--$M_V$ = 10.94.\label{fig6}}
\end{figure}

Figure 4 shows that the majority of stars concentrate in a relatively
narrow band of $A_V$ between 2.4 and 3.2 mag.  This means that behind
the clouds of the Great Cygnus Rift, located at about 700--900 pc, the
space is more or less free of dust, and this allows seeing stars up to 3
kpc with about the same extinction.  The selected 15 cluster members
(see Table 2 in Section 6) shown as open circles and triangles are
located at apparent distances from 1.35 to 1.80 kpc.  The extinction
$A_V$ values of these 15 stars are very different (from 2.5 to 3.8 mag),
the average value is 2.97 mag, which corresponds to $E_{B-V}$ = 1.03.

\section{THE INTRINSIC COLOR-MAGNITUDE\\ DIAGRAM}

Since the age of M29 is expected to be of the order of a few million
years, its color-magnitude diagram should contain ZAMS stars only in the
B star range.  The stars of lower masses should be on pre-main-sequence
evolutionary tracks.  The massive stars of early B subclasses are
expected to show evolutionary deviations from the ZAMS.

For the determination of the cluster age, we must know which stars are
the cluster members and what is their distance.  If the cluster has
sufficiently abundant population representing a broad range of color
indices, its distance can be easily estimated by fitting the ZAMS line
to the lower envelope of the main sequence in the intrinsic (dereddened)
color-magnitude diagram.  However, the population of M29 is rather poor,
and its main sequence in the color-magnitude diagram cannot be well
distinguished from the field B--A stars located in the foreground and
background.

The most natural way for determining the distance to M29 would be the
use of the group of six brightest stars without which the cluster would
be hardly recognizable.  Unfortunately, as it was mentioned above, these
stars are very problematic since their spectral types (especially
luminosity classes) determined by different authors show considerable
discrepancies, see Table 1.

For further analysis of the cluster, the spectral and luminosity classes
determined from our spectra by NRW were adopted.  The extinctions and
distances for these stars are given in the last two columns of Table 1.
The absolute magnitude calibration of MK types was taken from
\citet{Straizys1992}; the star 943 was considered as a ZAMS star, and
for the emission-line star 976 the luminosity class III was accepted.
The ratio $R_{YV}$ = $A_V$ / $E_{Y-V}$ = 3.83, corresponding to $R_{BV}$
= $A_V$ / $E_{B-V}$ = 2.87, was applied (see Section 3).

Table 1 shows that almost all early B-stars in M29, except HDE 229238,
exhibit similar distances, between 1.34 and 1.80 kpc, their mean value
is 1.54\,$\pm$\,0.15 kpc (standard deviation).  The errors of absolute
magnitude, $\pm 0.5$ mag, at 1.54 kpc give the distance
errors --0.32 kpc and +0.40 kpc.  Since individual distances of the
eight stars are contained within the distance error bars, we will
consider them being located at a mean distance of 1.54 kpc which
corresponds to the true distance modulus 10.94 mag.

Our next step is directed to estimate if more stars of cooler spectral
types, B3--B8, located in the color-magnitude diagram close to the ZAMS,
are present in the area within the same distance range (1.22--1.94 kpc).
In Figure 5 we plot the intrinsic diagram $V_0$ vs.  ($Y$--$V$)$_0$
for stars of spectral classes O, B and A in the
15$\arcmin$\,$\times$\,15$\arcmin$ area centered on the cluster.  The
ZAMS line shown is based on the \citet{Kazlauskas2006} table with
small corrections originating from the PARSEC isochrones up to
12\,$M_{\odot}$ \citep{Bressan2012}.  The ZAMS line was extrapolated
to larger masses (O8--B1 stars) with the older Padova isochrones
\citep{Bertelli1994, Girardi2002}.

The five brightest stars with $V$ between 8.9 and 9.3, defining the
visible cluster (after exclusion of the foreground star HD 194378) are
located in the left upper corner of the diagram, close to $Y$--$V$ = 0
and $V_0$ between 5.6 and 7.2 mag.  Most of them are slightly above the
ZAMS line.

In identifying possible ZAMS stars of the cluster with
$Y$--$V$\,$>$\,0.03 ($\geq$\,B1) we should be cautious since these stars
can be confused with the foreground and background field stars.  Among
the stars lying within 0.5 mag from the ZAMS and classified as
luminosity V or IV-V stars we have identified seven stars:  971 (B3),
1163 (B3), 897 (B5), 1339 (B7), 765 (B8), 775 (B8) and 1048 (B8).  These
stars (triangles in Figure 5) may also be cluster members, however, some
of them can be unresolved binaries deviating upward from their single
star positions.  Taking their $M_V$ values for ZAMS, we get their
distances from 1.37 kpc to 1.62 kpc, which within the errors are in
agreement with the above estimated distance of M29 at 1.54 kpc.
Luminosity classes of some of these stars from photometric
classification have been obtained as V-IV, but this is not an argument
against their attribution to ZAMS due to possible errors of colors and
$Q$-parameters used in the classification.  The cluster stars cooler
than B8 are not expected to be close to the ZAMS line since for an age
of the order of 5 Myr they should be pre-main-sequence objects.  Some of
them can be field stars.

The results reported above were based on Vilnius seven-color photometry.
We have also verified if the same results can follow taking the
observational data from the {\it B,V} photometric system.  Table 2
contains 14 stars measured in {\it UBV} by \citet{Massey1995}.  The
missing {\it BV} data for the star 1339 were taken from
\citet{Hoag1961}.  Their color excesses $E_{B-V}$ were calculated with
the intrinsic colors from \citet{Straizys1992}.  The ratio $R_{BV}$ =
2.87, determined in Section 3, has been applied.  The resulting
intrinsic color-magnitude diagram $V_0$ vs.  ($B$--$V$)$_0$ is shown in
Figure 6. Here the ZAMS line is defined by combining the Padova
isochrones for 1 and 10 Myr, and accepting the distance modulus
$V_0$--$M_V$ = 10.94.  It is evident that the positions of stars in this
diagram and the $V_0$ vs.  $(Y-V)_0$ diagram (Figure 5) are very
similar.  The stars between spectral classes B3 anf B8 lie close to the
ZAMS while O9--B2 stars deviate to higher luminosities.  Consequently,
the two independent photometric datasets confirm that the cluster
distance is really close to 1.5 kpc.

\section{THE PHYSICAL HR DIAGRAM}


\begin{table*}
\tabcolsep=5pt
\scriptsize
\caption{\baselineskip=9pt Probable members of the cluster M29. The luminosity classes Vz
correspond to ZAMS. In the last two columns, $p_{\rm S}$ and $p_{\rm D}$
mean the cluster membership probabilities (in percent) from \citet{Sanders1973} and
\citet{Dias2013}.}
\moveright2mm\vbox{
\begin{tabular}{rrlrrrrrrrrrrr}
\tablerule
{No.}        & {$V$}  & {Sp}&{$(Y-V)_0$}  &  {$E_{Y-V}$ }  &  {$A_V$}       &
 {$V_0$}      &  {$M_V$}       &  {$d$~kpc}    &
 {{\it BC}}   &  { $\log T_{\rm eff}$ } &
 {$\log L/L_{\odot}$ } &  { $p(\rm S)$  } &  {$p(\rm D)$}\\
\tablerule
 765 & 14.18 & b8\,Vz      & 0.12 &  0.68 & 2.60 & 11.58 &  0.90 & 1.37 & -0.55 &4.06 & 1.85 &    &  96 \\
 775 & 14.12 & b8\,Vz      & 0.12 &  0.64 & 2.45 & 11.67 &  0.90 & 1.42 & -0.55 &4.06 & 1.82 &    &  95 \\
 897 & 13.12 & b5\,Vz      & 0.09 &  0.67 & 2.57 & 10.55 & -0.20 & 1.41 & -1.30 &4.19 & 2.56 & 88 &  95 \\
 943 & 11.77 & B2\,V, SB   & 0.04 &  0.74 & 2.83 &  8.94 & -1.70 & 1.34 & -2.20 &4.36 & 3.57 &    &  95 \\
 971 & 12.21 & b3\,Vz      & 0.07 &  0.66 & 2.53 &  9.68 & -1.10 & 1.43 & -1.85 &4.29 & 3.13 & 77 &  96 \\
 976 &  9.22 & B0\,IIIe    & 0.00 &  0.89 & 3.41 &  5.81 & -5.00 & 1.45 & -2.95 &4.50 & 5.12 & 83 &  97 \\
1048 & 14.65 & b8\,Vz      & 0.12 &  0.74 & 2.83 & 11.82 &  0.90 & 1.53 & -0.55 &4.06 & 1.76 &    &  82 \\
1089 &  9.34 & B0\,III, SB & 0.00 &  0.80 & 3.06 &  6.28 & -4.10 & 1.80 & -2.95 &4.50 & 4.93 & 79 &  97 \\
1091 & 11.41 & b2\,V       & 0.04 &  0.81 & 3.10 &  8.31 & -2.60 & 1.52 & -2.20 &4.36 & 3.82 & 42 &  94 \\
1123 & 11.91 & B1.5\,V     & 0.03 &  1.00 & 3.83 &  8.08 & -2.05 & 1.64 & -2.40 &4.39 & 3.99 & 87 &  96 \\
1136 & 10.20 & B0.5\,V     & 0.01 &  0.78 & 3.00 &  7.20 & -2.80 & 1.55 & -2.95 &4.46 & 4.56 & 83 &  96 \\
1146 &  8.91 & O9\,III, SB & 0.00 &  0.79 & 3.03 &  5.88 & -5.20 & 1.64 & -3.15 &4.52 & 5.17 & 78 &  97 \\
1163 & 12.56 & b3\,Vz      & 0.07 &  0.72 & 2.76 &  9.80 & -1.10 & 1.51 & -1.85 &4.29 & 3.08 & 88 &  96 \\
1189 &  8.90 & B0\,III, SB & 0.00 &  0.85 & 3.26 &  5.64 & -5.00 & 1.34 & -2.95 &4.50 & 5.19 & 72 &  97 \\
1339 & 14.73 & b7\,Vz      & 0.11 &  0.83 & 3.18 & 11.55 &  0.50 & 1.62 & -0.80 &4.11 & 1.96 &    &  96 \\
\tablerule
\end{tabular}
}
\end{table*}

The possible members of M29 of spectral classes O9--B8 are listed in
Table 2. Figure 7 shows the $\log L/L_{\odot}$ vs.  $T_{\rm eff}$
diagram for these stars considering that they are located at 1.54 kpc
($V_0-M_V$ = 10.94).  The isochrones plotted for the ages 1, 5, 6 and 7
Myr are taken from the Padova database of stellar evolutionary tracks
and isochrones for the solar metallicity
\citep[$Z$=0.014,][]{Asplund2009}\footnote{~http://stev.oapd.inaf.it/cgi-bin/cmd}.
The isochrones and the ZAMS line correspond to the basic set of the
Padova isochrones up to the masses 120 $M_{\odot}$ \citep{Girardi2002}.
Luminosities of stars in solar units were calculated with the equation
\begin{eqnarray*}
\log L/L_{\odot}& = &  0.4 (M_{{\rm bol},\odot} - M_{{\rm bol},\star})=\\
                &   &0.4\,(4.72 - V + A_V + DM - BC),
\end{eqnarray*}

where $V$ is the apparent magnitude of the star, $M_{\rm bol,\star}$ is
its absolute bolometric magnitude, $M_{\rm bol,\odot}$ = 4.72 is the
bolometric absolute magnitude of the Sun, $BC$ is the bolometric
correction, and $DM$ is the true distance modulus of the cluster.  The
extinctions $A_V$ were determined with Eq.  (2).  The $DM$ in Figure 7
is taken as 10.94, which corresponds to a distance of 1.54 kpc.  The
effective temperatures and bolometric corrections were taken according
to their spectral types from Strai\v{z}ys (1992).  This $T_{\rm eff}$
scale for B-stars is close to those given by
\citet{Flower1996,Bessel1998} and \citet{Torres2010}.


\begin{figure}[!t]
\centerline{\includegraphics[width=75mm]{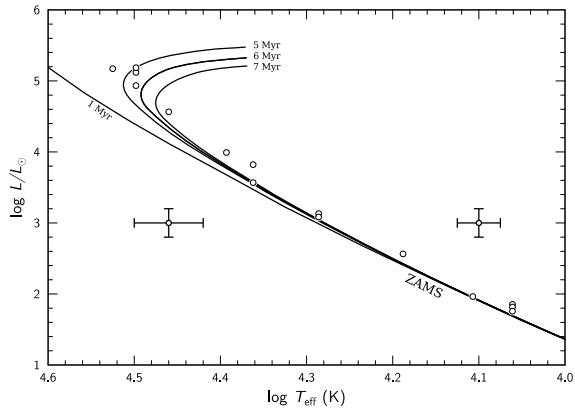}}
\caption{The effective temperature vs. luminosity diagram for the 15
probable members of the cluster and the Padova isochrones for the ages
1, 5, 6 and 7 Myr. The maximum (3$\sigma$) error crosses for the early
and late B subclasses are shown.\label{fig7}}
\end{figure}

It is evident that at a distance of 1.54 kpc the five most massive
members of M29 fit well with the 5 and 6 Myr isochrones.  Probably an
age of 5\,$\pm$\,1 Myr can be accepted.  The fainter stars deviate from
the ZAMS less than $\Delta \log L/L_{\odot}$ = 0.2, which corresponds to
0.5 mag in absolute magnitude, thus their positions within the errors do
not contradict cluster membership.  Their deviations from the
ZAMS can be explained by the errors of (a) spectral or photometric
classification, (b) calibration of MK types in terms of temperatures and
BCs, (c) distance modulus determination, and (d) the isochrones
themselves.  Unresolved duplicity also cannot be excluded.  Figure 7
shows the error bars for the early and late B-subclasses.  The
errors in temperatures correspond to $\pm$\,0.5 spectral subclass, and
the errors in luminosities correspond to $\pm$\,0.5 mag.

\section{DISCUSSION}

It seems that our investigation has solved the long standing problem of
the distance and age of the cluster M29. The main reason preventing
determination of the cluster distance earlier is related to
insufficient accuracy of the luminosity classifications of its brightest
stars.

The luminosity class errors may originate from different reasons.  The
simplest source of errors can be the differences in quality and
resolving power between the spectra of the program and the comparison
stars which can make the list of luminosity criteria inapplicable.  The
early classifications, in which the photographic method has been used,
could suffer from the sensitivity variations across the plates, i.e.,
insufficient ``flat-fielding''.  \citet{Boeche2004} point also to a
possible effect on the intensities of the compared lines which fall on
different Echelle orders.

Therefore it is important to have additional criteria for determination
of cluster membership.  One such criterion might be the proper motions
of stars in the cluster area.  At present, proper motions of great
numbers of stars are availale in the astrometric catalogs, such as UCAC,
PPMXL and others.  In the area of M29, proper motions determined by
\citet{Sanders1973} are available, however their limiting magnitude is
too shallow.  As we mentioned in the Introduction, in the direction
along the Local spiral arm, proper motions are small, and their errors
frequently are larger than the proper motions themselves.  However,
Table 2 shows that all the selected probable members have high
membership probabilities given by \citet{Sanders1973} and by
\citet{Dias2013}, the latter determined from proper motions given in the
UCAC4 catalog.  Unfortunately, many stars, which are definitely
foreground and background stars, also have the membership probabilities
close to 100.  This means that proper motions are not sufficient to have
trustworthy memberships.  Radial velocities are much more informative
\citep[see NGC 6819,][]{Tabetha2009}.  According to \citet{Liu1989} and
\citet{Boeche2004}, the heliocentric velocity of M29 is --16.4 and
--16.9 km/s, respectively.  Boeche et al. find that seven stars from
Table 2 (943, 976, 1089, 1123, 1136, 1146 and 1189) are radial velocity
members.  For the remaining eight stars radial velocities have not been
measured.

In the present paper, the cluster members were identified mainly from
their visible crowding and from spectroscopic and photometric distances.
As the result, 15 member stars of M29 within the
15$\arcmin$\,$\times$\,15$\arcmin$ area centered on the cluster were
selected.  Many previous investigations have also used spectroscopic and
photometric distances, but usually they showed a much larger dispersion
and even bimodal distribution, originating from the errors in luminosity
classes of the most luminous stars.  Also, in most of the previous
papers the normal value of the ratio $R = A_V / E_{B-V}$ has been
applied.  We have found that in the direction of M29 the ratio $R$ is
lower than the standard value, and this resulted in a larger distance of
the cluster stars by a factor of about 1.15.  The range of the
extinctions for cluster members is in a good agreement with the
surrounding field stars (Figure 4).

Considering the cluster distance, the decisive factor was the
application of the new MK spectral types of the brightest stars
determined with the criteria developed by \citet{Walborn1990}
and \citet{Sota2011} and applied to classification of O--B stars of
M29 by NRW himself.  The results are in perfect agreement with the
photometric data in the Vilnius and {\it B,V} systems and the Padova
isochrones.

It may seem that a shortcoming in our distance and age determination is
the neglect of spectroscopic binarity of the four O9--B0 stars
discovered by \citet{Liu1989} and \citet{Boeche2004}.  If their
components were of the same spectral type, the single stars should be
fainter by 0.75 mag and more distant by a factor of 1.41.  However,
since the spectral lines of the secondaries are not observable,
differences in magnitudes between the components should be much larger.
In this respect it is interesting to note that the exclusion of the four
spectroscopic binaries from the distance determination does not change
the mean distance itself -- it remains 1.54 kpc.  A small vertical shift
of the binaries down in the HR diagram also does not change the estimate
of the age since isochrones in the range of O9--B1 spectral types run
almost vertically.

The intrinsic color-magnitude diagram in Figure 5 shows that there are a
number of stars in the B3--A7 range lying above the ZAMS up to 2.5 mag.
Some of them can be field stars and others can be binary stars of the
cluster.  Distances and reddenings of some of these stars with
luminosities IV and III allow them to be the evolved cluster members.
This could imply that star formation in the cluster began about 50
Myr ago and is still on-going.  While this is not impossible, we note
the absence of a prestellar dust and gas cloud in the cluster vicinity.
Most probably, the main star forming activity in the cluster took place
about 5--6 Myr ago, and after that the massive luminous O--B stars
dissipated the remains of the parent cloud.

If the low-mass YSOs have been formed at the same time as the massive
ones, they should be present in the area of M29. We estimated the
apparent magnitudes of possible T Tauri stars of spectral type K in M29
comparing it with the cluster NGC 2264 which is of similar age and
contains pre-main-sequence stars both of high and low masses.  In this
cluster PMS stars of early K subclasses appear at $V$ = 13.5 mag
\citep{Park2000}.  Since the distance to NGC 2264 is 760 pc
\citep{Dahm2008}, the difference of the true distance moduli of both
clusters is 1.54 mag.  The mean extinctions $A_V$ are 0.2 mag for NGC
2264 and $\sim$\,3 mag for M29. Consequently the difference of the
apparent distance moduli of both clusters is about 4.3 mag.  The
conclusion is that in M29 YSOs of class II and spectral class K are
expected to be fainter than $V$ about 17.8 mag or $J$ about 16.4 mag.
This means that in M29 YSOs of Class II should be invisible in 2MASS
photometry (see the limiting magnitudes in \citet{Skrutskie2006}) but
they should be easily detectable in the UKIDSS survey \citep{Lucas2008}.
In the UKIDSS $J$--$H$ vs.  $H$--$K$ diagram of the area tens of sources
above the intrinsic T Tauri line are seen, but their identification as
YSOs of M29 is problematic.  A special study, using the data from the
UKIDSS, Spitzer, WISE, IPHAS and other surveys, is essential.

The M29 cluster is usually considered to be a member of the Cyg OB1
association, along with the clusters IC 4996, Berkeley 86 and Berkeley
87; see the recent review by \citet{Reipurth2008}.  Other than the
clusters, the association contains tens of O--B3 stars and cooler
supergiants scattered in an area of 4$\degr$\,$\times$\,3.5$\degr$
\citep{Humphreys1978,Garmany1992}.  The distance to this association is
still controversial but is usually considered to be 1.3--1.5 kpc.  The
distance to the cluster M29 determined in this investigation is close to
the largest of these distances.  According to \citet{Hanson2003}, the
massive association Cyg OB2 located $\sim$\,3$\degr$ north of M29 can be
also located at the same distance.  In this case the centers of Cyg OB2
and M29 are separated by $\sim$\,80 pc.  However, both star groups
probably are not physically related since their movements are somewhat
different:  the heliocentric radial velocity is --10.3 km/s for Cyg OB2
\citep{Kiminki2007} and --16.9 km/s for M29 \citep{Boeche2004}.

\section{RESULTS AND CONCLUSIONS}

This investigation is based on the results of seven-color photometry and
two-dimensional classification of 260 stars down to $V$\,$\approx$\,19
mag in a 15$\arcmin$\,$\times$\,15$\arcmin$ area centered on the cluster
M29 (NGC 6913), 2MASS {\it JHK}$_s$ photometry, and the blue spectra for
the nine brightest stars of M29 obtained at the Asiago Observatory.  The
spectra were classified in spectral and luminosity classes with the
criteria described by \citet{Walborn1990} and \citet{Sota2011}.
The following results were obtained:

(1) The ratio of color excesses is found to be $E_{J-H}/E_{H-K_s}$ =
2.044\,$\pm$\,0.080.  This means that the interstellar extinction law in
the near infrared spectral range is close to normal.  However, the ratio
$R = A_V / E_{B-V}$, found from the ratios $E_{V-i} / E_{B-V}$ for the
$i$ = $J$, $H$ and $K_s$ passbands, is close to 2.87, i.e. somewhat
lower than the normal value $R = 3.15$.  In the violet and ultraviolet
spectrum down to 300 nm the normal interstellar extinction law is valid
again.

(2) The distance to M29, 1.54\,$\pm$\,0.15 kpc, is determined as the
average of distances of eight stars of spectral types O9-B2 classified
from the Asiago spectra.  Two stars from the seven brightest
ones are found to be field stars:  HDE 229238 is a background B0.5
supergiant and HD 194378 is a foreground F star.

(3) In the intrinsic color-magnitude diagram of O-B-A stars we selected
seven more stars of spectral classes B3--B8 lying close to the ZAMS
which may be also considered as the cluster members.  The cooler stars
were not considered since they should be still in the pre-main-sequence
stage of evolution.

(4) The 15 probable members of the cluster of spectral classes O9--B8,
plotted on the $\log L/L_{\odot}$ vs.  $\log T_{\rm eff}$ diagram
together with the isochrones from the Padova database, give the age of
the cluster 5\,$\pm$\,1 Myr. The extinction $A_V$ is variable across the
cluster (from 2.5 to 3.8 mag), the average value being 2.97 mag, which
corresponds to $E_{B-V}$ = 1.03.

(5) The distances to the clouds of the Great Cygnus Rift are estimated
to be in the range of 700--900 pc.

{ACKNOWLEDGMENTS}.  The use of the Simbad, WEBDA, ADS, SkyView,
2MASS and UKIDSS databases is acknowledged.  We are grateful to Wilton
S. Dias for membership probabilities in the M29 area presented before
publication.  The paper is partly supported by the Research Council of
Lithuania, grant No.  MIP-061/2013.

\end{document}